\documentclass[aip, apl, amsmath, amssymb, reprint, floatfix] {revtex4-1}

\usepackage{graphicx}
\usepackage{dcolumn}
\usepackage{bm}

\usepackage[utf8]{inputenc}
\usepackage[T1]{fontenc}
\usepackage{mathptmx}
\usepackage{etoolbox}
\usepackage{hyperref}
\usepackage{threeparttable} 
\usepackage{tabularx}
\usepackage{booktabs}
\usepackage{adjustbox}
\usepackage{float}
\makeatletter
\def\@email#1#2{%
 \endgroup
 \patchcmd{\titleblock@produce}
  {\frontmatter@RRAPformat}
  {\frontmatter@RRAPformat{\produce@RRAP{*#1\href{mailto:#2}{#2}}}\frontmatter@RRAPformat}
  {}{}
}%
\makeatother
\begin{document}

\preprint{AIP/123-QED}

\title[APL]{Hydrophilic direct bonding of (100) diamond
and a deposited SiO$_2$ substrates}
\author{Tianyin Chen}
\affiliation{Department of Quantum and Computer Engineering, Delft University of Technology, Netherlands}

\author{Jeffrel Hermias}
\affiliation{Department of Quantum and Computer Engineering, Delft University of Technology, Netherlands}
\author{Salahuddin Nur}%
\email{r.ishihara@tudelft.nl}
\affiliation{Department of Quantum and Computer Engineering, Delft University of Technology, Netherlands}

\author{Ryoichi Ishihara}
\homepage{r.ishihara@tudelft.nl}
\affiliation{Department of Quantum and Computer Engineering, Delft University of Technology, Netherlands}
\affiliation{QuTech, Delft University of Technology, Netherlands}

\date{\today}

\begin{abstract}
Diamond has emerged as a leading material for solid-state spin quantum systems and extreme environment electronics. However, a major limitation is that most diamond devices and structures are fabricated using bulk diamond plates. The absence of a suitable diamond-on-insulator (DOI) substrate hinders the advanced nanofabrication of diamond quantum and electronic devices, posing a significant roadblock to large-scale, on-chip diamond quantum photonics and electronics systems. In this work, we demonstrate the direct bonding of (100) single-crystal (SC) diamond plates to PECVD-grown SiO$_2$/Si substrates at low temperatures and atmospheric conditions. The surfaces of the SiO$_2$ and diamond plates are then activated using oxygen plasma and piranha solution, respectively. Bonding occurs when the substrates are brought into contact with water in between and annealed at 200$^{\circ}$C under atmospheric conditions, resulting in a DOI substrate. We systematically studied the influence of piranha solution treatment time and diamond surface roughness on the shear strength of the bonded substrate, devising an optimal bonding process that achieves a high yield rate of 90$\%$ and a maximum shear strength of 9.6 MPa. X-ray photoelectron spectroscopy (XPS) was used for quantitative analysis of the surface chemicals at the bonding interface. It appears that the amount of -OH bindings increases with the initial roughness of the diamond, facilitating the strong bonding with the SiO$_2$. This direct bonding method will pave the way for scalable manufacturing of diamond nanophotonic devices and enable large-scale integration of diamond quantum and electronic systems.

\end{abstract}
\maketitle


Diamond is a promising material for quantum applications\cite{DOHERTY20131, 10.5555/3312180, tvDSar2021, MaxRuf_Network2021}. Spins in diamond color-centers such as NV center have shown significant potential in quantum technologies\cite{DOHERTY20131, 10.5555/3312180, Bradley_network_2022}. Their long spin coherence times\cite{bar-gillSolidstateElectronicSpin2013, abobeihOnesecondCoherenceSingle2018}, optical readout capabilities, high operational temperatures\cite{PhysRevLett.119.253601}, and suitability for nanofabrication techniques and nanophotonic structures make them highly attractive for quantum applications. Recently, tin-vacancy (SnV) centers\cite{PhysRevLett.119.253601, PhysRevX.14.041008, PhysRevX.11.041041} in diamond showed their reduced sensitivity to environmental noise and compatibility with nanofabricated devices\cite{Rugar_WG_2020, 10.1063/5.0051675, PhysRevLett.133.023603, Chen:24}, enhancing their versatility for large-scale quantum computers\cite{ishihara3DIntegrationTechnology2021} and communication \cite{MaxRuf_Network2021, PhysRevX.11.041041,Chen:24}. Diamond is also an exceptional material for power and extreme environment electronics\cite{PEREZ2020108154, 10.1088/1361-6463/ab4eab, nano14050460, liang_direct_2024, jne2040032, khanna_harsh2023, 9798392} due to its superior electronic and thermal properties\cite{10.1007/s10853-023-08232-w}, including high electrical breakdown field strength, exceptional carrier mobility, and excellent thermal conductivity\cite{PEREZ2020108154, 10.1088/1361-6463/ab4eab, nano14050460, jne2040032}. These characteristics make diamond an ideal candidate for devices requiring efficient heat dissipation and high-power handling capabilities\cite{PEREZ2020108154, 10.1088/1361-6463/ab4eab, nano14050460, jne2040032, khanna_harsh2023, 9798392, 10.1007/s10853-023-08232-w}. Beyond its electronic advantages, diamond also possesses outstanding optical properties\cite{opt_eng_dia}, such as a high refractive index and wide bandgap\cite{hiscocksDiamondWaveguidesFabricated2008}, along with superior mechanical properties like a high Young's modulus\cite{10.1063/1.3683544}.

Despite these promising attributes, scalable integration of diamond-based systems remains elusive, especially in quantum technologies. This challenge is primarily due to the lack of appropriate substrates. Epitaxial growth or deposition of high-quality single-crystal diamond is extremely challenging \cite{KASU2016317, 10.1038/srep44462}. Moreover, commercially available single-crystal diamond substrates are small—typically less than a centimeter or even just a few millimeters due to the HTHP process\cite{10.1134/s1063782618110271, 10.1039/c9ra06126f} —and most diamond quantum devices to date are fabricated from these bulk diamond substrates \cite{Burek2012, Khanaliloo_2015, TOROS2020107839, Rugar_WG_2020, 10.1063/5.0051675, PhysRevLett.133.023603, Chen:24}. In quantum photonic applications, two critical factors are essential. First, the quality of the interface is crucial, as any damage, trapped charges, or poor interface quality can significantly degrade the optical properties of color center qubits, reduce coherence times, and impair overall system performance\cite{PhysRevX.9.031052, Rodgers2021, PhysRevX.13.011042, kumar2024}. Second, enhancing the optical performance of diamond color center qubits often requires integrating photonic devices. This necessitates the use of diamond-on-insulator (DOI) substrates. For high-quality device manufacturing and scalable integration in both quantum technologies and power electronics, DOI substrates would be a key component. The method for producing DOI structures is to use wafer or substrate bonding technologies to attach diamond plates to different substrate materials.
\begin{table*}[!ht]
  \caption{Diamond direct bonding methods review}\label{methodsreview}
  \centering
  \begin{threeparttable}
    \setlength{\tabcolsep}{1mm} 
    \small 
    \adjustbox{max width=\textwidth}{ 
      \begin{tabular}{@{} p{3cm} p{4cm} p{4cm} p{3cm} p{2cm} p{2cm} p{3cm} p{1.5cm} @{}} 
        \toprule
        \textbf{Bonding Type} & \textbf{Substrates bonded} & \textbf{Diamond Surface treatment} & \textbf{Process conditions (Roughness, Pressure, Temperature)} & \textbf{Bonded Area (up to)} & \textbf{Bonding Strength} & \textbf{Remark/critical issues for Quantum applications} & \textbf{Ref.} \\
        \midrule
        \raggedright Hydrophilic & \raggedright Diamond (111), Various semiconductor substrates (InP, Si) and SiO$_2$/Si (thermal or native) & \raggedright Oxidizing solutions, such as H\textsubscript{2}SO\textsubscript{4}/H\textsubscript{2}O\textsubscript{2} and NH\textsubscript{3}/H\textsubscript{2}O\textsubscript{2} mixtures, \\$\sim 75^\circ$C, Atm & \raggedright $S_a$ < 0.5 nm, \\ 0 to 1 MPa Load, $\sim 200^\circ$ to $250^\circ$C & \raggedright $5 \times 5$ mm & \raggedright $\sim 10$ to $35$ MPa & \raggedright Amorphous  layer $\sim 3$ to 5 nm, \\Not suitable for photonics  & \cite{matsumaeHydrophilicDirectBonding2020, MATSUMAE202024, 9159285, matsumaeHydrophilicDirectBonding2019,  fukumotoHeterogeneousDirectBonding2020, fukumotoHeterogeneousDirectBonding2020, fukumoto_direct_2021, liang_direct_2024, matsumae_low-temperature_2021, liang_room_2022, matsumae_simple_2023, matsumae_low-temperature_2020}\\
        \hline
        \raggedright Hydrophilic & \raggedright Diamond (100), Si & \raggedright Oxidizing solutions, H\textsubscript{2}SO\textsubscript{4}/H\textsubscript{2}O\textsubscript{2} and NH\textsubscript{3}/H\textsubscript{2}O\textsubscript{2} mixtures, \\$\sim 75^\circ$C, Atm & \raggedright $S_a$ $\sim 0.1$ to 0.2 nm, \\Pressure $\sim 10$ MPa, \\$\sim 200^\circ$ to $250^\circ$C & \raggedright $5 \times 5$ mm & \raggedright 1.7 MPa & \raggedright Amorphous layer $\sim 3$ to 5 nm, \\weaker bond strength, \\Not suitable for photonics & \cite{matsumaeLowtemperatureDirectBonding2021, okita_optimizing_2023} \\
        \hline
        \raggedright Surface activated bonding (SAB) & \raggedright Diamond (100), Sapphire & \raggedright Ar beam irradiation, \\High vac $<1\times 10^{-5}$ Pa & \raggedright $S_a$ < 0.2 nm, \\20 MPa, \\RT & \raggedright $4 \times 4$ mm & \raggedright $\geq 14$ MPa & \raggedright 300 nm amorphous layer, \\Diffusion of bonding elements, \\Poor interface for quantum applications & \cite{miyatake_surface-activated_2023}\\
        \hline
        \raggedright Adhesive & \raggedright Diamond substrates, Si or SiO\textsubscript{2}/Si substrates & \raggedright Boiling Piranha solution & \raggedright $S_a$ < 0.4 nm or 2 nm, \\HSQ adhesion, \\Pressure 0 to 105 kPa, \\500 to 600$^\circ$C, \\Membrane synthesis (smart-cut) and transfer & \raggedright  $ 200 \times 200 \,\mu$m \\to $1 \times 1$ mm & \raggedright Strong, but not measured & \raggedright Negligible amorphous layer, \\HSQ background fluorescence, \\Poor scalability and integration & \cite{tao_single-crystal_2014, kuruma_telecommunication-wavelength_2021, jung_reproducible_2016, guo_tunable_2021, guo_microwave-based_2023} \\
        \hline
        \raggedright Plasma activation based (PAB) & \raggedright Diamond to various substrates (Si, fused silica, Sapphire, thermal oxide, lithium niobate) & \raggedright  O\textsubscript{2} plasma ashing (O\textsubscript{2} flow - 200 sccm and RF power - 600 W for 150 s) for hydrophilic surface & \raggedright $S_a$ < 1 nm, \\Smart-cut and transfer,  \\Heating up to $170^\circ$C, \\No pressure, \\$550^\circ$C anneal & \raggedright $ 200 \times 200 \, \mu$m & \raggedright Strong, but not measured & \raggedright Sub-nm interfacial layer, nm uniformity, \\DOI Film $\geq 10$ nm, \\Poor scalability and integration & \cite{guoDirectbondedDiamondMembranes2023, ding_high-q_2024} \\
        \hline
        \raggedright Hydrophilic & \raggedright Diamond (100), PECVD SiO\textsubscript{2}/Si & \raggedright Oxidizing solutions - H\textsubscript{2}SO\textsubscript{4}/H\textsubscript{2}O\textsubscript{2} mixtures, $\sim 75^\circ$C, Atm & \raggedright $S_a$ $\sim$ 1.5 to 5 nm, \\No pressure, \\$200^\circ$C & \raggedright $4.5 \times 4.5$ mm & \raggedright $\sim 9$ MPa & \raggedright Negligible intermediate layer, \\Suitable for quantum \& photonics, Scalable &  In this work  \\

        \bottomrule 
        
      \end{tabular}
    }
  \end{threeparttable}
\end{table*}

Bonding methods are usually categorized based on the use of intermediate materials/layers to attach the substrates: non-direct and direct bonding. A variety of bonding techniques, including Surface Activated Bonding (SAB)\cite{liangRealizationDirectBonding2017, miyatake_surface-activated_2023, liangAnnealingEffectSurfaceactivated2019}, , Atomic Diffusion Bonding (ADB) \cite{matsumae_room-temperature_2018}, Compression Bonding \cite{delmas_thermal_2024}, Metallic Bonding \cite{wang_room_2023}, Adhesive Bonding \cite{heupel_fabrication_2020, bleiker_adhesive_2017, tao_single-crystal_2014, kuruma_telecommunication-wavelength_2021, jung_reproducible_2016, guo_tunable_2021, guo_microwave-based_2023}, Plasma Activated Bonding (PAB) \cite{guoDirectbondedDiamondMembranes2023, ding_high-q_2024}, Fusion Bonding \cite{yushinStudyFusionBonding2002, piracha_scalable_2016}, and Hydrophilic Direct Bonding \cite{matsumaeHydrophilicDirectBonding2019, matsumaeHydrophilicLowtemperatureDirect2020, matsumaeHydrophilicDirectBonding2020, matsumaeDirectBondingDiamond2020, matsumaeLowtemperatureDirectBonding2021, okita_optimizing_2023, fukumotoHeterogeneousDirectBonding2020, fukumotoHeterogeneousDirectBonding2020, fukumoto_direct_2021, liang_direct_2024, matsumae_low-temperature_2021, liang_room_2022, matsumae_simple_2023, matsumae_low-temperature_2020}, have been employed to attach diamond plates to other substrates \cite{zhao_progress_2024}. A summary of direct bonding techniques used for diamond substrates is provided in Table \ref{methodsreview}. Most of these approaches can lead to significant interface issues, such as amorphization, metallization, or defects, which make the bonded substrates unsuitable for complex, large-scale 3D integration processes \cite{ishihara3DIntegrationTechnology2021} and for quantum applications. The direct bonding methods are the most promising for producing DOI-type substrates necessary for on-chip diamond-based quantum systems. Matsumae et al., showed that OH-terminated diamond (111) and Si or SiO$_2$/Si substrates are chemically bonded through a dehydration reaction\cite{matsumaeHydrophilicDirectBonding2020, MATSUMAE202024, 9159285, matsumaeHydrophilicDirectBonding2019}. However, the previous studies have been limited to extremely smooth (< 0.5 nm) diamond (111) substrates and thermally grown or native SiO$_2$ interface layers. It should be noted that (100) sc-diamond is widely available and suitable, especially for group IV color-centers in diamond such as SnV center. Also, (111) sc-diamond has even smaller size due to cutting along (111) plane of the (100) diamond plates. Furthermore, the thermally grown SiO$_2$ requires high-temperature processes and hence limits the applications, such as heterogeneous integration of diamond on temperature-sensitive materials. In this work, we have achieved a hydrophilic direct bonding of diamond (100) substrates onto SiO$_2$ layers grown by PECVD on silicon wafers, by optimizing the surface roughness of diamond. Additionally, we have examined the chemical composition of diamond surfaces after Piranha treatment and analyzed the influence of treatment conditions on the shear strength of the bonded interfaces.


The experimental process involves the following steps: First, a (100)-oriented diamond substrate with different roughness undergoes an immersion in  Piranha solution (H$_2$SO$_4$ and H$_2$O$_2$ in a 3:1 ratio) at 75$^\circ$C for various treatment times to achieve OH termination. After treatment, the diamond substrate is rinsed in deionized water for 5–10 minutes. SiO$_2$/Si wafers are prepared separately, a 300 nm layer of silicon dioxide is deposited onto the Si wafer using a PECVD reactor (Novellus Concept 1) at a temperature of 400$^\circ$C. After the deposition, the SiO$_2$/Si wafers are diced into substrates, which are OH-terminated using $\text{O}_2$ plasma at 1000 W for 5 minutes. This OH-termination of SiO$_2$ through plasma treatment facilitates chemical bonding between the diamond and SiO$_2$/Si substrates in the presence of water. Next, the diamond substrate is placed atop the $\text{SiO}_2$/Si substrate in the presence of water without needing any external pressure. The assembled substrates are stored under atmospheric conditions (20$^\circ$C and 40\% relative humidity) for 3 days to remove excess water molecules. Finally, the specimens are annealed at 200$^\circ$ for 24 hours to initiate the reaction \cite{matsumaeHydrophilicDirectBonding2019}, which describes the bond formation process between diamond and $\text{SiO}_2$ interface:

\begin{equation}
\mathrm{C}-\mathrm{OH}+\mathrm{HO}-\mathrm{Si} \rightarrow \mathrm{C}-\mathrm{O}-\mathrm{Si}+\mathrm{H}_2 \mathrm{O}
\end{equation}

\begin{figure} [htbp]
\includegraphics[width=0.8\columnwidth]{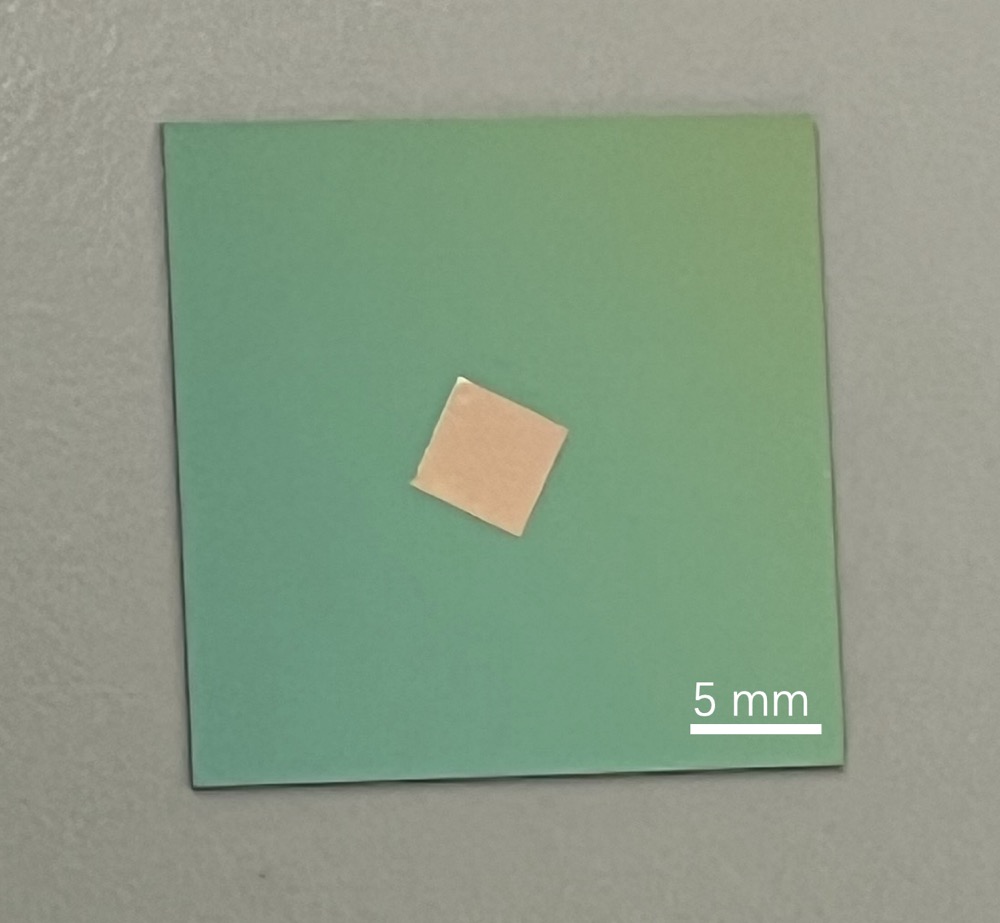}
\caption{\label{bondedsampel}A 4mm $\times 4$mm diamond substrate is bonded onto a 25mm $\times 25 $mm SiO$_2$/Si substrate. There is no air gap in the bonding interface.}
\end{figure}

\begin{figure} [htbp]
\includegraphics[width=\columnwidth]{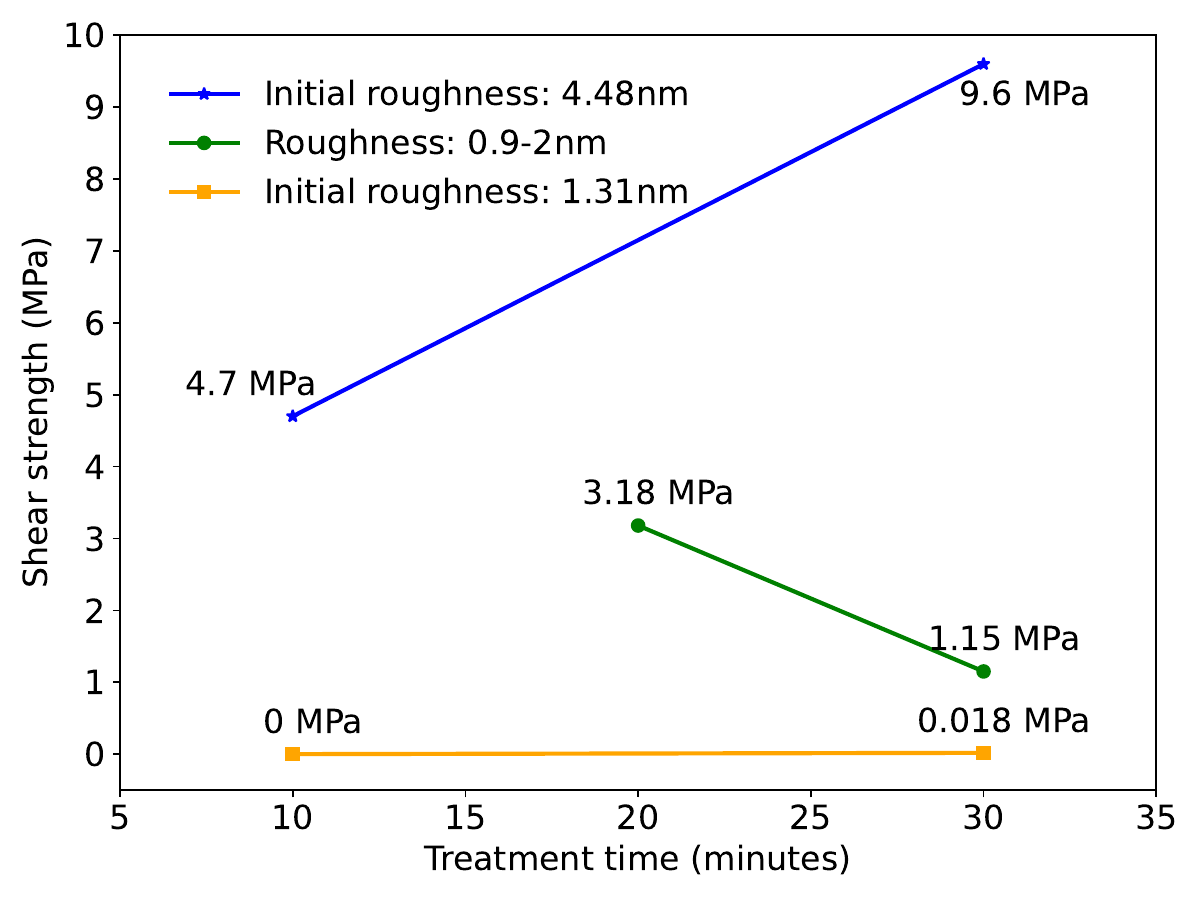}
\caption{\label{shear} Shear strength as a function of Piranha treatment time. Three sources of diamond substrates with different initial roughness were used. The diamond substrate with highest initial roughness (4.48nm, indicated by solid blue datapoints) shows an increasing trend as treatment time increases. However, with smaller initial roughness, diamond substrate shows an decreasing trend (roughness range 0.9-2nm, solid green datapoints) or inability to be bonded (1.31nm, solid yellow datapoints).}
\end{figure}

Figure \ref{bondedsampel} presents a photograph of a (100) diamond substrate bonded to a SiO$_2$/Si substrate after Piranha treatment at 75$^{\circ}$C for 30 minutes. In the image, the blue substrate is SiO$_2$/Si, and the transparent substrate is diamond (4 mm $\times$ 4 mm, thickness 50 $\mu$m) and there is no visible fringes. This result demonstrates that diamond cleaned with proper Piranha solution can be effectively bonded to a SiO$_2$/Si substrate without visible air gaps. Figure\ref{shear} shows the shear strength of diamond substrates bonded to 300 nm SiO\textsubscript{2}/Si. We have utilized three types of diamond substrates with different initial roughnesses—4.48 nm, 2 nm, and 1.31 nm. Surface roughness has been evaluated using an atomic force microscope (AFM, Bruker). After successful bonding, bonding strength has been measured using a Nordson DAGE 4000Plus die shear tester. The testing procedure adheres to the guidelines specified in the MIL-STD-883 standard. The rough diamonds with an initial roughness of 4.48 nm exhibit the highest shear strength (4.7 MPa and 9.6 MPa) compared to the other two experimental groups, and there is an increasing trend with longer treatment time. In contrast, the smooth diamonds with an initial roughness of 1.31 nm are not able to be bonded even with the long treatment time. This contrasts with findings in previous works\cite{matsumaeHydrophilicDirectBonding2020, MATSUMAE202024, 9159285, matsumaeHydrophilicDirectBonding2019}, which show that only very smooth diamond (111) substrates can be bonded. Interestingly, the intermediate group with a roughness of 2 nm exhibits a decreasing trend in the moderate shear strength as the treatment time increases. This may be because when the diamond roughness reaches a certain threshold, roughness becomes the dominant factor affecting bonding, favoring rougher surfaces. Conversely, when the diamond roughness is below that threshold, other factors, such as surface chemistry and activation, play more significant roles.

\begin{figure} [htbp]
\includegraphics[width=\columnwidth]{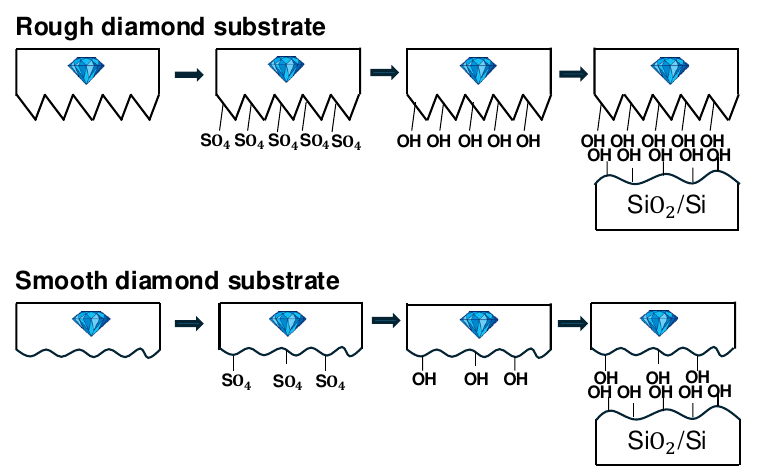} 
\caption{\label{Schemetic} 
Schematic for possible diamond bonding mechanism. Rough surfaces provide more areas to be covered with -OH groups which improves the success of bonding.}
\end{figure}

Chemical composition of the substrates have been identified as critical factors for successful bonding. The chemical composition of the treated diamond surface has been investigated using X-ray photoelectron spectroscopy (XPS). Analyses have been carried out using a PHI TFA XPS spectrometer (Physical Electronics Inc.), equipped with an X-ray Al K$\alpha$ monochromatic source ($hv=1486.7$ eV). The vacuum during XPS analysis has been maintained at approximately $10^{-9}$ mbar. During measurements, the analyzed area has a diameter of 0.4 mm and a corresponding depth of analysis in the range of 3–5 nm. High-resolution narrow multiplex scans of C1s, O1s, S2p3, and Si2p peaks have been collected with pass energies of 23.5 eV and a resolution of 0.2 eV at a take-off angle of $45^\circ$C. The acquired spectra have been processed using MultiPak v8.0 (Physical Electronics Inc.). Figure\ref{Schemetic} schematically illustrates the possible mechanism of diamond direct bonding.


\begin{figure}[htbp]
\centering
  \includegraphics[width=0.94\columnwidth]{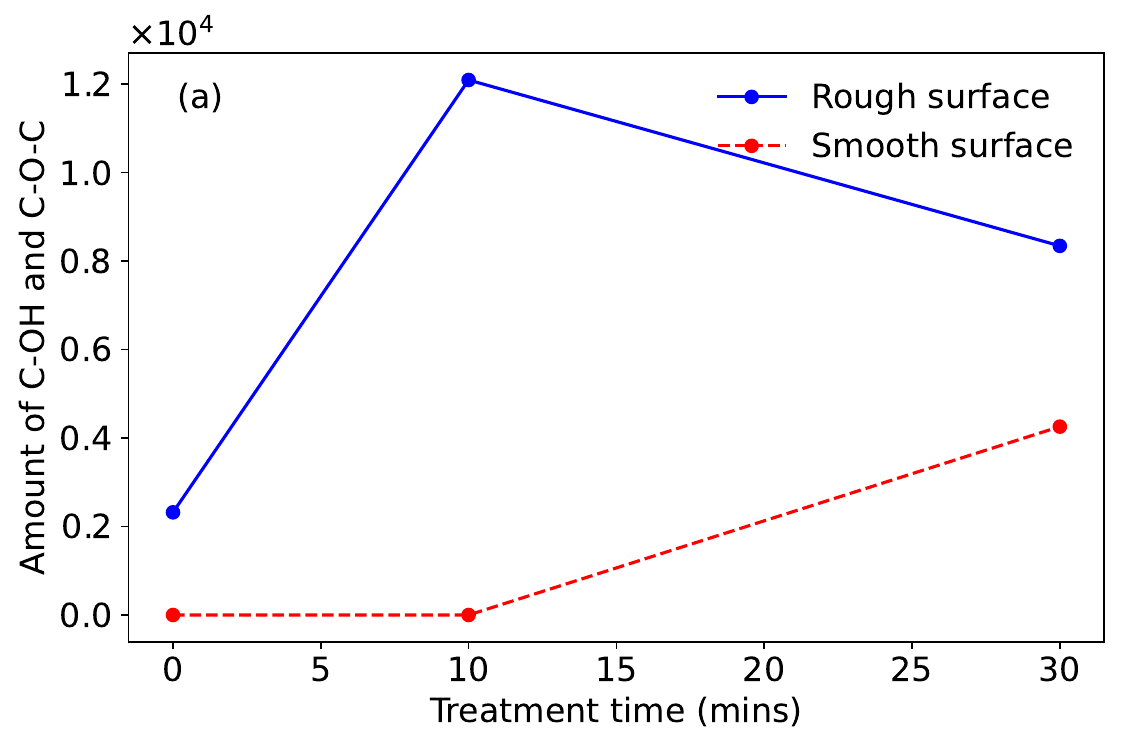}
  \vspace{5mm}  
  \includegraphics[width=0.94\columnwidth]{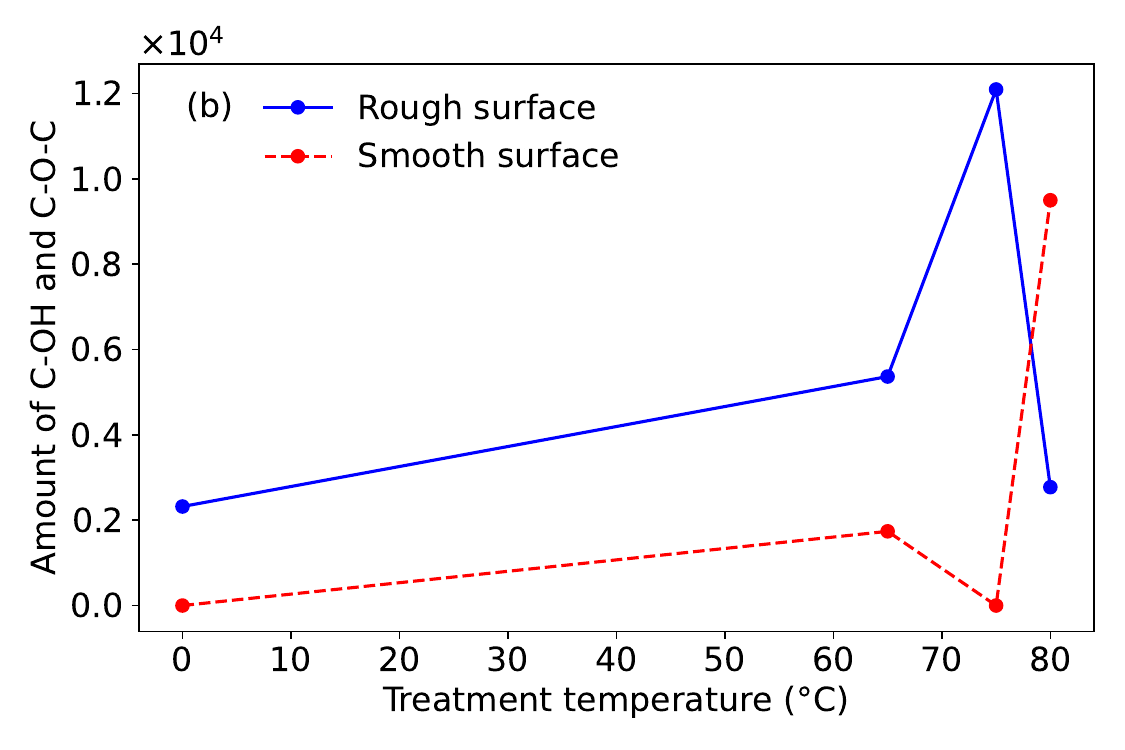}
  \vspace{5mm}
  \includegraphics[width=0.94\columnwidth]{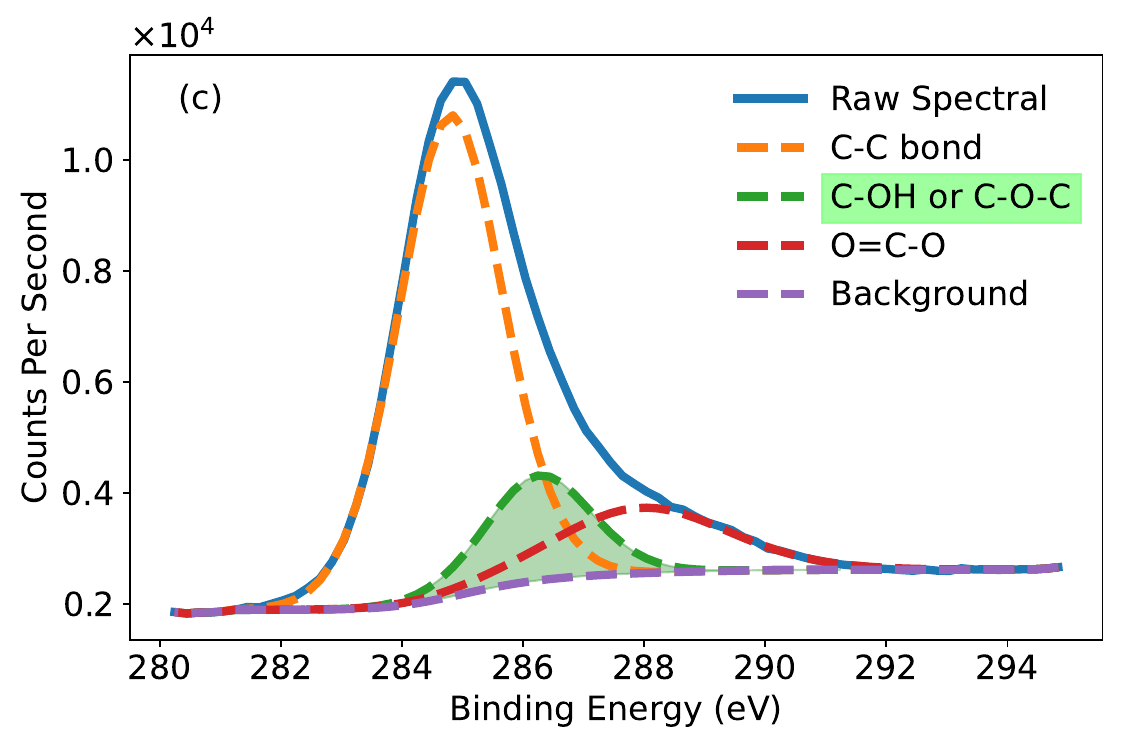}
\caption{\label{xps}
Amount of surface C-OH groups as a function of Piranha treatment time and temperature, and detailed XPS spectrum of a rough diamond surface treated with 10 mins of Piranha under 75$^\circ$C. (a) Treatment temperature is fixed and time duration is the variable. (b) Treatment time is fixed and temperature is the variable. (c) XPS raw scan was deconvoluted into 3 peaks, where the green area in the spectrum represents the amount of C-OH or C-O-C groups.}
\end{figure}

Our XPS results, plotted in Figure\ref{xps}, support this mechanism by illustrating the variation of C–OH groups as a function of treatment time and temperature. We observe that rough surfaces tend to exhibit a higher amount of C–OH groups compared to smooth surfaces. The inset spectra in Figure\ref{xps}(a) show the fitted peaks of the C1s region from an XPS spectrum of a diamond sample treated with Piranha solution for 10 minutes at 75$^\circ$C. The fitted peaks are attributed to C–C, O=C–O, and C–OH or C–O–C bonds (note that C–OH and C–O–C cannot be differentiated by XPS\cite{fukumotoHeterogeneousDirectBonding2020}). Additionally, Figure \ref{xps}(a) includes a line chart illustrating the variation in the quantity of C–OH or C–O–C bonds with different treatment durations, comparing rough and smooth surfaces of diamond substrates. These substrates have one side polished to a roughness less than 2 nm. Before wet treatment, the rough surface shows the presence of C–OH or C–O–C bonds, while the smooth surface has no signal in this region. This signal can be attributed to the native C–O–C bonds existing on the rough surface. After Piranha treatment, diamond surfaces are terminated with newly generated –OH groups. To mitigate the influence of surface roughness on the detected surface chemical groups, we focus on the data from the smooth sides, as they have similar roughness values of around 2 nm. Analysis of the smooth surface data reveals that the amount of –OH groups increases with longer treatment time. Additionally, the temperature of the Piranha solution significantly affects the generation of –OH groups. In the temperature range of 65–80$^\circ$C, treatment at 80$^\circ$C appears to be more effective in promoting the generation of –OH groups. Therefore, Piranha treatment time and temperature are two dominant factors in surface activation. Moreover, rough surfaces tend to be more readily terminated with –OH groups. Figure \ref{xps}(b) shows the variation of Piranha-terminated –OH groups as a function of treatment temperature, indicating that a 75$^\circ$C Piranha bath is preferred as the optimal condition considering both effectiveness and practical considerations. Figure \ref{xps}(c) shows a deconvoluted spectrum example of rough diamond surface treated in Piranha solution for 10 mins under 75$^\circ$C. C-C, C-OH (or C-O-C), O=C-O peaks show up in the spectrum, and C-OH groups provides the source of direct bonding. To validate our conclusions, it is good to compare XPS measurements with bonding strength measurements. The amount of –OH groups directly affects the strength of the bonding. Therefore, aligning the results of the –OH group quantification with bonding strength under different treatment conditions provides further confirmation of our findings.

Our combined XPS and bonding strength measurements suggest that both surface roughness and the generation of –OH groups are critical in determining the bonding strength. While higher roughness facilitates the formation of more –OH groups, leading to stronger bonds, overly smooth surfaces may lack sufficient reactive sites for effective bonding. Before treatment, the rough surfaces of the diamond are covered with native C–C and C–O–C groups, while the smooth surfaces are only covered with native C–C groups. When the diamond surface is treated with Piranha solution, hydroxyl (C–OH) groups are generated via dehydration reactions. On the one hand, the rough surface has more available areas to be terminated with –OH groups due to its higher surface area. On the other hand, the native C–O–C groups, which have the same valence state as C–OH, provide an additional source for C–OH generation. This leads to a significantly higher generation of –OH groups on rough surfaces compared to smooth surfaces. In the final step, the chemical reaction between a large amount of the –OH groups for the initially rough surface on the diamond and SiO$_2$ surfaces forms a strong bond between the two substrates.

In conclusion, We have demonstrated a hydrophilic direct bonding of (100) single-crystal diamond plates—with thicknesses of 50 $\mu$m and 500 $\mu$m—to PECVD-grown SiO$_2$/Si substrates under low-temperature and atmospheric conditions, creating diamond-on-insulator (DOI) substrates suitable for advanced quantum and electronic devices. Our findings reveal that surface roughness in the range of 2 nm to 5 nm and chemical composition are critical for strong bonding; specifically, relatively rough diamond surfaces (with roughness around 4.48 nm) generate more hydroxyl (–OH) groups after Piranha treatment due to their higher surface area and the presence of native C–O–C groups, leading to stronger chemical bonds at the interface. By optimizing Piranha treatment conditions and surface roughness within this range, we achieved a 90\% bonding yield and a maximum shear strength of 9.6 MPa. This direct bonding method addresses the need for scalable DOI substrates, paving the way for large-scale integration of diamond-based systems. Future work will focus on refining this technique to produce photonic-grade DOI substrates and exploring its applicability to larger substrates and other insulating materials, thereby advancing diamond nanophotonic devices and on-chip quantum technologies.

\begin{acknowledgments}

We gratefully acknowledge support from the joint research program “Modular quantum computers” by Fujitsu Limited and Delft University of Technology, co-funded by the Netherlands Enterprise Agency under project number PPS2007.

\end{acknowledgments}

\section*{Data Availability Statement}

The data that support the findings of this study are available from the corresponding authors upon reasonable request.

\nocite{*}
\bibliography{aipsamp} 

\end{document}